\documentclass[nofootinbib,aps]{revtex4}
\usepackage{amstext,amsmath,amssymb}
\usepackage[dvips]{graphicx}
\usepackage{latexsym}
\usepackage{psfrag}

\newcommand{\Ref}[1]{(\ref{#1})}

\newcommand{\N}{\mathbb{N}}

\newcommand{\C}{\mathbb{C}}

\def\be{\begin{equation}}
\def\ee{\end{equation}}
\def\bes{\begin{eqnarray}}
\def\ees{\end{eqnarray}}
\def\arr{\rightarrow}
\newcommand{\dr}{\rightarrow}

\def\la{\langle}
\def\ra{\rangle}
\def\f{\frac}

\newcommand{\lalg}[1]{\mathfrak{#1}}  
\newcommand{\SU}{\mathrm{SU}}
\newcommand{\SO}{\mathrm{SO}}
\newcommand{\U}{\mathrm{U}}
\newcommand{\su}{\lalg{su}}
\renewcommand{\sl}{\lalg{sl}}
\renewcommand{\u}{\lalg{u}}

\newcommand{\osp}{\lalg{osp}}

\renewcommand{\v}{\overrightarrow}
\def\dag{^\dagger}

\def\upa{\uparrow}
\def\downa{\downarrow}

\def\cc{{\cal C}}
\def\mm{{\cal M}}
\def\dd{{\cal D}}
\def\ss{{\cal S}}

\begin{document}

\title{Reconstructing Quantum Geometry from Quantum Information:\\
Spin Networks as Harmonic Oscillators}
\author{{\bf Florian Girelli}\footnote{fgirelli@perimeterinstitute.ca}, {\bf Etera R. Livine}\footnote{elivine@perimeterinstitute.ca}}
\affiliation{Perimeter Institute, 31 Caroline Street North, Waterloo, ON, Canada N2L 2Y5}
\date{\today}
\begin{abstract}
\begin{center}{\ ABSTRACT}\end{center}

Loop Quantum Gravity defines the quantum states of space geometry as spin networks and describes their evolution in time. We reformulate spin networks in terms of harmonic oscillators and show how the holographic degrees of freedom of the theory are described as matrix models. This allow us to make a link with non-commutative geometry and to look at the issue of the semi-classical limit of LQG from a new perspective. This work is thought as part of a bigger project of describing quantum geometry in quantum information terms.

\end{abstract}

\maketitle

\tableofcontents

\section{Introduction}

Loop Quantum Gravity (LQG) is a candidate theory for quantum gravity (for reviews, see \cite{lqg}). It achieves a canonical quantization of general relativity: it describes the quantum states of space geometry and their evolution in "time". The states are the so-called spin network states, which can be roughly understood as quantized discretized manifold. The theory derives a discrete spectrum for geometrical operators (such as the areas and volumes) and therefore implements the notion of quanta of space(time).

There three main issues which need to be addressed in loop quantum gravity. A first one is the problem of the {\it semi-classical limit}: we would like to recover general relativity (and quantum field theory) as low energy theories. More precisely, the theme is to truly understand the geometry of a spin network background, to understand how the quantum geometry of spin networks will describe a smooth spacetime manifold at some coarse-grained level. We must identify diffeomorphisms acting on the discrete quantum geometry defined by a spin network background (maps from itself to itself). We must develop the notion of spacetime points in this quantum context and how to localize objects and systems in such a quantum spacetime.

A second issue is to solve the dynamics of the theory. We are supposed to implement and solve the Hamiltonian constraint in order to identify the physical states of the theory. Then we should somehow recover a notion of time and evolution through some gauge fixing.

The third issue is to find physically relevant observables. In this diffeomorphism invariant context, a natural choice is the relational observables, which can be understood as the physical correlations between two systems. If one of these systems can be identified with a reference frame (when its state space approximates a classical symmetry group), then one has constructed an observable describing the position of the other system with respect to that reference system. In our quantum geometry set-up, we are particularly interested in $\SO(3)$ reference frame, which would allow to define a direction in space, and $\U(1)$ reference frame, which would define a (quantum) clock.

A possibly relevant point of view is provided by quantum information. It appears that many mathematical tools and physical questions are actually shared by loop quantum gravity and the quantum information field, and it could be helpful to develop an explicit bridge between these two research areas. This defines a project to reformulate quantum geometry in terms of entropy, entanglement and quantum reference frames. The present work is thought as part of this bigger project. We seek to understand the geometry of spin networks at the kinematical level. We reformulate spin networks in terms of harmonic oscillators and we show how the boundary degrees of freedom in LQG are described as matrix models. More precisely, we show that the space of $\SU(2)$ intertwiners with $N$ legs carries a $\u(N)$ structure. This allows a fresh look on the semi-classicality issue from the point of view of non-commutative geometry. We also describe how spin network can be understood as quantum circuits, i.e. that the quantum gravity kinematical states can be interpreted as dynamical processes. This particular idea of spin networks arising as a natural tool in describing quantum computations was already pushed forward by Marzuoli and Rasetti \cite{marzuoli}.
In a last section, we explain how quantum reference frames should allow us to localize systems on a spin network background and end with the idea that a notion of distance could be constructed on a spin network can be constructed in terms of entanglement between parts of that spin network.

\section{Spin Networks, Surface States and Harmonic Oscillators}

Loop Quantum Gravity describes the quantum states of 3d geometry as {\it spin networks}, which roughly describe one-dimensional quantized excitations of the gravitational field. Spin networks are graph whose edges are dressed up with $\SU(2)$ representations and whose vertices are $\SU(2)$ invariant tensors intertwining between the representations associated to the edges linked to a particular vertex. More precisely, we use the irreducible finite-dimensional representations of $\SU(2)$, which are labeled by a half-integer usually called {\it spin} and noted as $j$. The geometrical interpretation of a spin network is as a discretized 3d manifold. Indeed the area of a surface  and the volume of a space region become operators, which turn out to be diagonal in the spin network basis: the spin $j$'s label the area eigenvectors while the intertwiners define the volume eigenvalues. This way, the vertices of a spin network represent chunks of volume, and the edge representing the link between two chunks of volume is dual to the surface separating the two regions. The area of an elementary surface intersecting the underlying spin network at only an edge, which is labeled by a spin $j$, can be derived from the quantization of the area operator and is defined up to regularization issues. Noting $l_P$ the Planck length, the area of a spin $j$ is usually assumed to the square root of the Casimir operator, $a(j)\equiv \sqrt{j(j+1)}l_P^2$, but other reasonable candidates are the simpler choice $a(j)\equiv j l_P^2$ or the shifted choice $a(j)\equiv (j+1/2)l_P^2$.

\medskip

Let us now describe the state of a surface in the LQG framework. A generic surface is considered as made of a certain (finite) number of elementary patches or surfaces. An elementary surface is defined as one corresponding to a single intersection with a spin network. It is thus described by a $\SU(2)$ irreducible representation labeled by a spin $j$. The spin $j$ gives the area of the elementary surface while a generic vector $|j m\ra$ of the representation Hilbert space $V^j$ can be understood as the quantized version of the (bi)vector normal to the surface.

We propose to use a different basis for the Hilbert space corresponding to an elementary surface exploiting the presentation of $\SU(2)$ as a system of two harmonic oscillators. Instead of (quantum) vectors labeled by their "norm" $j$ and their "direction" $m$, we will have states labeled by the energy level of the two harmonic oscillators. We hope that such a change of point of view would lead to some further insight in the dynamics of spin networks in LQG since harmonic oscillator systems are better understood than any other systems in theoretical physics.

Explicitly, let us start with two harmonic oscillators:
$$
[a,a\dag]=[b,b\dag]=1, \qquad
[a,b]=0.
$$
Then we define the following operators:
\be
J_z=\f{1}{2}(a\dag a-b\dag b), \qquad J_+=a\dag b, \qquad J_-=J_+\dag=ab\dag,
\qquad E=\f{1}{2}(a\dag a+b\dag b).
\ee
The operators $J$'s define a $\su(2)$ algebra while the total energy $E$ commutes with the $J$'s and is thus a Casimir operator for $\su(2)$:
\be
[E,\cdot]=0, \qquad [J_z,J_\pm]=\pm J_\pm, \qquad [J_+,J_-]=2J_z.
\ee
The spin representations at fixed $j$'s are given by fixing the total energy $E$. Then diagonalising the two operators $E$ and $J_z$, we obtain the simple correspondence between the $\SU(2)$ usual basis $|j m\ra$ and the basis defined by the energy levels $|n_a n_b\ra$ of the two oscillators:
\be
j=\f{1}{2}(n_a+n_b),\quad
m=\f{1}{2}(n_a-n_b), \qquad
E\,|j m\ra=j|j m\ra, \quad
J_z\,|j m\ra=m|j m\ra.
\ee
Let us remark that the usual quadratic Casimir operator of the $\su(2)$ algebra is given by:
\be
{\cal C}\equiv \vec{J}^2= J_z^2+\f{1}{2}\left(J_+J_-+J_-J_+\right)=E(E+1).
\ee
Here, in a sense, considering the harmonic oscillators allows us to take the square root of the quadratic Casimir.

We have introduced these harmonic oscillators as a mathematical tool. Physically, one is tempted to interpret the oscillators $a$ and $b$ are left and right modes moving along the edge of the spin network puncturing the elementary surface. The total energy $E$ then defines the area of the surface, while the difference of energies defines the orientation and direction of the surface. From this point of view, the elementary surface can be considered as an elementary (chiral) screen with information flowing through it from left to right and right to left, in such a way that the information flow itself defines the geometry (state) of the surface. This might help implementing the ideas on the holographic principle for loop quantum gravity presented in \cite{screen}.

Let us point out that we are here using the area spectrum $j\times l_P^2$, defined by the energy operator $E$, instead of the standard $\sqrt{j(j+1)}\times l_P^2$ usually used in LQG. By considering, the $1/2$ additional term in the energy of the harmonic oscillator (defining the energy by $2E=\{a,a\dag\}=2a\dag a +1$), then we would obtain the $(j+1/2)\times l_P^2$ area spectrum. Due to regularization ambiguities in Loop Quantum Gravity, there is no mathematical reason to select a particular area spectrum over another, and the present harmonic oscillator presentation favors the use of the simpler spectrum $j\times l_P^2$.

After having described surface states, we will study in the next section the structures associated to volumes in quantum geometry, i.e. $\su(2)$ intertwiners, and we will show how the harmonic oscillator presentation of $\su(2)$ allow for a simple description of the volume states. 

\section{Intertwiners and $\SU(N)$ structure}

Let us consider a single vertex (of a spin network) with $N$ edges linked to it. Labeling the edges with $\su(2)$ spins $j_1,..,j_N$, the vertex is described by an intertwiner between the $N$ representations, i.e. an $\SU(2)$-invariant tensor $V^{j_1}\otimes..\otimes V^{j_N}\arr\C$. The intertwiner defines the (quantum) volume corresponding to the vertex. Indeed it contains information on how the elementary surfaces, dual to the edges, are combined together to form a surface boundary of the space region dual to the vertex. 
Mathematically, $j_i$ gives the area of the $i$th elementary surface and corresponds to the $\SU(2)$ Casimir operator $\v{J_{(i)}}.\v{J_{(i)}}$ of the representation $V^{j_i}$ while the intertwiner contains information on how these representations are intertwined and provides the values of the mixed Casimir operators $\v{J_{(i)}}.\v{J_{(j)}}$ (which defines the angle between the elementary surfaces $i$ and $j$). Now, from the point of view of the boundary surface, we are considering a 2-sphere with a single vertex inside from which $N$ links are leaving and puncturing the sphere at $N$ points which represent the $N$ elementary surfaces the surface is made of.

Here, we are interested by characterizing the algebra of $\SU(2)$-invariant observables associated to a vertex with $N$ edges, or equivalently a surface made of $N$ elementary patches, at fixed total area. More precisely, we now allow the spins $j_1,..,j_N$  to vary while we keep the total area ${\cal A}=\sum_i j_i$ fixed, and we look for the algebra of operators commuting with a global $\SU(2)$ gauge transformation (on all elementary surfaces)  and with the operator defining the total area.
In this picture, not only the intertwiner is fuzzy and undetermined but also the surface itself is fuzzily defined, so that all the geometric quantities are treated on the same footing.

\begin{figure}[t]
\begin{center}
\psfrag{a}{$a$}
\psfrag{b}{$b$}
\psfrag{j}{$j$}
\includegraphics[width=8cm]{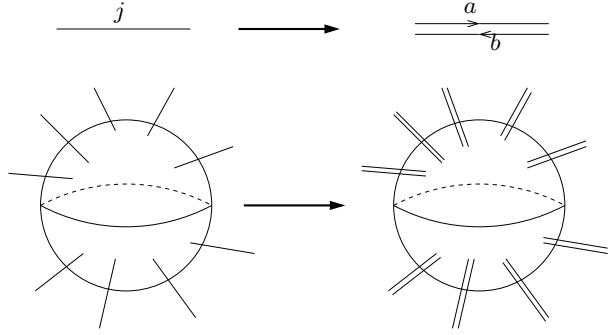}
\end{center}
\caption{Using the presentation of $\SU(2)$ in terms of two harmonic oscillators, we replace all lines of the spin network by double lines symbolizing the harmonic oscillators $a$ and $b$. Then, considering an intertwiner with $N$ legs, which can be interpreted as dual to a $\ss^2$ surface punctured by $N$ spin networks edges, we look for $\SU(2)$ invariant states in the tensor product of the $2\times N$ harmonic oscillators.}
\end{figure}

Having $N$ elementary surfaces, we start with $N$ copies of the $\su(2)$ algebra $\v{J_{(i)}}$ and equivalently with $N$ pairs of harmonic oscillators $a_i,b_i$. The total area of the surface is defined as the sum of the area of all the elementary patches, and therefore as the total energy of the $2N$ harmonic oscillators:
\be
{\cal A}\equiv E\equiv \sum_{i=1}^N E_i=\f{1}{2}\sum_i(a_i^\dagger a_i+b_i^\dagger b_i).
\ee
Global $\SU(2)$ gauge transformations on the full surface is defined as the simultaneous action of all the local $\SU(2)$'s:
\be
J_z=\sum_i^N J_z^{(i)}, \quad J_\pm=\sum_i^N J_\pm^{(i)}.
\ee
We now seek to identify the algebra of operators invariant under global $\SU(2)$ rotations and leaving the total area/energy $E$ invariant. Considering the action of the operators $a_i,a_i^\dagger, b_i,b_i^\dagger$, it is straightforward to see that this invariant algebra is generated by the local area operators acting each puncture:
$$
E_i=\f{1}{2}\left(a_i^\dagger a_i+b_i^\dagger b_i\right),
$$
and by cross operators acting (non-locally) on each pair of punctures:
$$
E_{ij}=\f{1}{2}\left(a_i^\dagger a_j+a_j^\dagger a_i+b_i^\dagger b_j+b_j^\dagger b_i\right),
$$
$$
F_{ij}=\f{i}{2}\left(a_i^\dagger a_j-a_j^\dagger a_i+b_i^\dagger b_j-b_j^\dagger b_i\right).
$$
This way, we obtain $N+2N(N-1)/2=N^2$ operators. It is straightforward to check that the $E_i,E_{ij},F_{ij}$ are all Hermitian operators and form a $\lalg{u}(N)$ Lie algebra. Then, quotienting by the trivially invariant operator $E=\sum_i E_i$, we get that the invariant algebra is $\lalg{su}(N)$.

More explicitly, introducing the operators $G_{ij}=E_{ij}+iF_{ij}$ and $G\dag_{ij}=E_{ij}-iF_{ij}=G_{ji}$, we have the following commutation relations:
\be
[G_{ij},G\dag_{ij}]=2(E_j-E_i),
\quad
[G_{ij},E_i]=+\f{1}{2}G_{ij},
\quad
[G_{ij},E_j]=-\f{1}{2}G\dag_{ij},
\quad
[G_{ij},G_{ki}]=G_{kj}
\label{commutator}
\ee

\medskip

Now, it is rather natural to obtain $\U(N)$ as the group of operators leaving the area invariant, and it can be thought as the action of the {\it diffeomorphisms} moving $N$ punctures on the surface of a 2-sphere. 
When increasing the number of punctures, we refine the structure of the surface, building it out of more and more points or patches, and we refine these discrete diffeomorphisms defined by $\U(N)$ towards the continuum limit. There are actually many works explaining how $\U(\infty)$ can be thought as the diffeomorphism group of a 2-sphere (see references in \cite{taylor}).

Nevertheless, one could object that spin networks describe (3d) diffeomorphism invariant states of geometry, so how can we talk of diffeomorphisms acting non-trivially on a spin network. In fact, we are dealing with diffeomorphisms moving the stuff (degrees of freedom) living on the spin networks. Just as diffeomorphisms on a manifold (which is abstractly a diffeomorphism invariant object) is a map from the manifold to itself moving the manifold points around, one can introduce a similar notion of diffeomorphisms living on a spin network moving around the degrees of freedom living on the edges and vertices of the spin networks.

\medskip

One should compare the $\U(N)$ structure, that we derive here by allowing the punctures to take any spin value, with the standard case when all the punctures have a fixed spin, usually all at spin 1/2.
In that case, it is possible to decompose the tensor product $(V^{1/2})^{\otimes N}$ into the direct sum of tensor products of a $\SU(2)$ representation with spin running from 0 to $N/2$ and a representation of the permutation group of $N$ objects. Formally, it reads:
\be
(V^{1/2})^{\otimes N}=\bigoplus_{k=0}^{N/2}\left(
V^k \otimes R^{(k)}({\cal S}_N)
\right),
\ee
where $R^{(k)}({\cal S}_N)$ is an irreducible representation of the permutation group ${\cal S}_N$.
This is usually referred as Schur duality (see for example \cite{goodman}). That tensor product decomposition shows that the intertwiner space between $N$ punctures of fixed spin $1/2$ provides an irreducible representation of ${\cal S}_N$. In our generalized analysis, by allowing the punctures to have any possible spin, we extend the discrete group ${\cal S}_N$ to the Lie group $\U(N)$.

\medskip


\subsection{The Structure of Intertwiner States}

While the operators $E_i$ probe the area of a single puncture, the mixed operators $G_{ij},G\dag_{ij}$ probe the structure of the intertwiner between the punctures. Indeed considering two punctures $i,j$, the intertwiner between the two representations is defined through the values of the Casimir operators $\v{J_{(i)}}.\v{J_{(j)}}$ which can be expressed in terms of the $E$ and $G$ operators:
\be
\v{J_{(i)}}.\v{J_{(j)}}
=\f{1}{2}G\dag_{ij}G_{ij}-E_iE_j-E_i
=\f{1}{2}G\dag_{ij}G_{ij}-\left(E_i+\f{1}{2}\right)\left(E_j+\f{1}{2}\right)+\f{1}{2}(E_j-E_i)+\f{1}{4}.
\ee
Therefore, the operators $E,G$ contain the same information as the $\v{J_{(i)}}.\v{J_{(j)}}$ operators and allow to describe all the $\SU(2)$ gauge invariant operators and the full intertwiner space.

Moreover, this formula allows us to extract the explicit spectrum of the positive operator $G\dag_{ij}G_{ij}$. Considering that
$$
\v{J_{(i)}}.\v{J_{(j)}}=\f{1}{2}\left(
(\v{J_{(i)}}+\v{J_{(j)}})^2-\v{J_{(i)}}^2-\v{J_{(j)}}^2
\right),
$$ 
and having fixed the representations corresponding the punctures $i$ and $j$ to  $\v{J_{(i)}}^2=a(a+1)$ and
$\v{J_{(j)}}^2=b(b+1)$ with $a,b\in\N/2$, it is well-known that the allowed values of $(\v{J_{(i)}}+\v{J_{(j)}})^2$ are $c(c+1)$, with $|b-a|\le c\le (a+b)$. In this basis $|a,b,c\ra$, the operator $G\dag_{ij}G_{ij}$ is diagonal and its values are:
$$
G\dag_{ij}G_{ij}\,|a,b,c\ra= 
c(c+1)-(b-a)(b-a+1)\,|a,b,c\ra.
$$
Therefore assuming $a\le b$ and introducing the positive level $n=c-(b-a)$ which maximal value is $2a$, the eigenvalues of $G\dag_{ij}G_{ij}$  are $2n(b-a)+n(n+1)$.
The first term $n\times 2(b-a)$ can be identified as a harmonic oscillator contribution with $n$ the number of quanta since \Ref{commutator} gives the commutator $[G_{ij},G\dag_{ij}]=2(b-a)$ as soon as we diagonalize the operators $E_i,E_j$. The extra-term $n(n+1)$ can then be interpreted as coming from the $\SU(2)$ structure. In the end, we can write the spectrum of $G\dag_{ij}G_{ij}$ as depending on two numbers, the energy level (of the mixing of the two punctures) $n\in\N$ and the energy difference (of the two punctures) $k=2(b-a)\in\N$:
\be
\lambda(n,k)=n(n+k+1).
\ee

\medskip

Let us now discuss the structure of the intertwiner space as seen from the $\U(N)$ point of view. First let us recall that an intertwiner is a $\SU(2)$ invariant tensor between representations of $\SU(2)$. More precisely, choosing the spin representations $j_1,..,j_N$ living at the $N$ punctures, an intertwiner is a map $\otimes_i V^{j_i}\arr \C$ which is invariant under global $\SU(2)$ rotations. Let us call ${\cal I}_{j_1,..,j_N}$ the Hilbert space of such intertwiners. Actually, in our framework , we consider arbitrary spins at each puncture. Thus noting $W=\oplus_j V^{j}$ the (reducible) $\SU(2)$ representation containing all possible spins, we have been looking at intertwiners $W^{\otimes N}\arr\C$. Let us call ${\cal I}$ the (Hilbert) space of such intertwiners. As the operators $\v{J_{(i)}}^2$ and $\v{J_{(i)}}.\v{J_{(j)}}$ are invariant under global $\SU(2)$ rotations, they define operators on ${\cal I}$. Moreover as the operators $\v{J_{(i)}}^2$ commute with each other, they can be be simultaneously diagonalized, which means that:
$$
{\cal I}=\bigoplus_{j_1,..,j_N}{\cal I}_{j_1,..,j_N}.
$$
The interesting structure of the intertwiner space comes from the fact that the cross operators $\v{J_{(i)}}.\v{J_{(j)}}$ do not commute and can not be simultaneously diagonalized. Moreover, we have constructed above the $\SU(2)$ invariant operators $E_i, G_{ij}$, which are thus operators on ${\cal I}$, who generate all $\SU(2)$ invariant operators i.e all operators acting on ${\cal I}$. As these operators form a $\u(N)$ algebra, the Hilbert space of intertwiners ${\cal I}$ carries a $\U(N)$ representation.

More precisely, picking a given representation of $\U(N)$, a vector in the representation space specifies an intertwiner. In more details, the operators $E_i$ form the Cartan subalgebra of $\u(N)$ while the cross operators $G_{ij}$ are the roots of the Lie algebra. The irreducible (unitary) representations of $\u(N)$ are highest weight representations. The weights give the eigenvalues of the operators $E_i$, which we note $\alpha_1,..,\alpha_N$. So a representation is defined by the maximal values of the $\alpha$'s ${\cal R}(\alpha^{max}_i)$ and a basis or the representation space can be roughly labeled as $|{\cal R}(\alpha^{max}_i), \alpha_1,..,\alpha_N\ra$. Now interpreting such a vector as an intertwiner, we point out that the weights $\alpha_i$ are actually the spin labeling the $\SU(2)$ representations: $j_i\equiv \alpha_i$.  So $|{\cal R}(\alpha^{max}_i), \{\alpha_i\}_i\ra$ defines an intertwiner in ${\cal I}_{\alpha_i}$. Then on such a intertwiner state, one can compute the expectation values and correlations of the $G_{ij}$ operators and thus of the invariant operators $\v{J_{(i)}}.\v{J_{(j)}}$. This will be investigated in details in \cite{cohstates} at least in the case $N=4$ (first non-trivial case).

\medskip

An important issue in (Loop) Quantum Gravity is to build semi-classical states of geometry. Here we develop a notion of coherent intertwiners, which we hope to be useful in the study of semi-classicality. We define coherent intertwiner states as minimizing the uncertainty relations on the invariant operators $\v{J_{(i)}}.\v{J_{(j)}}$.
More explicitly, while the $J_{(i)}^2$'s define the area associated to each puncture (or more precisely dual to each puncture), the $\v{J_{(i)}}.\v{J_{(j)}}$'s can be interpreted as describing the area attached to internal surfaces. The issue is that these internal area operators do not commute. Indeed it is easy to check that:
\be
\left[\v{J_{(i)}}.\v{J_{(j)}}\, ,\, \v{J_{(i)}}.\v{J_{(k)}}\right]=i\epsilon_{abc}J_{(i)}^aJ_{(j)}^bJ_{(k)}^c\,\ne0.
\label{vol}
\ee
This commutator can actually be considered as defining a volume\footnote{For $N=4$ punctures, the 4-valent intertwine is dual to a tetrahedron. The operator $U\equiv \epsilon_{abc}J_{(1)}^aJ_{(2)}^bJ_{(3)}^c$ defines the square volume operator for this quantum tetrahedron \cite{barbieri}. It is easy to check that
$$
U=-i\left[\v{J_{(i)}}.\v{J_{(j)}}\, ,\, \v{J_{(i)}}.\v{J_{(k)}}\right]
=\f{i}{4}\left(
G_{ki}G_{jk}G_{ij}-G_{ji}G_{kj}G_{ik}
\right)
=\f{i}{4}\left(
G_{ij}G_{jk}G_{ki}-G_{ik}G_{kj}G_{ji}
\right),
$$
in our formalism. More details on the properties of this operator will be explained in \cite{cohstates}.}. However, the point we would like to insist upon here is that the Lie algebra generated by the commutators of the $\v{J_{(i)}}.\v{J_{(j)}}$'s doesn't close and generates operators of higher and higher order in the initial $J$'s. Now if one would like to write coherent intertwiner states, or semi-classical states, one would need in principle to minimize the uncertainties relations corresponding to each of these commutation relations\footnotemark, i.e. a infinite tower of uncertainty relations. On the other hand, our formalism gives us access to the operators $E,G$ from which we can reconstruct all the $\SU(2)$ invariant operators in the $J$'s. These operators $E,G$ form a closed Lie algebra identified as $\u(N)$ and we do not need to investigate the detailed structure of the tower of commutators of the $J$ operators. One can then use all the already developed formalism of coherent states on Lie groups \cite{perelomov}. Roughly, coherent states are given up to a unitary transformation by the highest weight states of the irreducible (unitary) representations of $\u(N)$ with $\alpha_i=\alpha^{max}_i$. The details of the analysis of these coherent states will be carried elsewhere \cite{cohstates}.

\footnotetext{If we start with the commutation relation between Hermitian operators [A,B]=iC, the corresponding uncertainty relation is:
$$(\delta A). (\delta B)\ge \f{1}{2}\la C\ra.$$}

One further advantage to study directly the $\u(N)$ states is that we are dealing at once with all possible values of the puncture areas, instead of the standard naive approach of minimizing the uncertainty relation \Ref{vol} case by case for each set of values of the boundary areas $J_{(i)}^2$. 

\subsection{Fuzzy Sphere, Fuzzy Surface and Classical Limit}

Another very interesting approach to Quantum Gravity, besides Loop Quantum Gravity (LQG) is Non-Commutative Geometry  (NCG), either as a framework describing an effective Quantum Gravity theory (Deformed Special Relativity, non-commutative field theories,...) or defining quantum spacetime as a full non-commutative spacetime (Connes approach). There is hardly an explicit link between LQG and NCG (except in 3d gravity). Identifying a relationship between the two would be of crucial importance as it would provide more tools to study issues of Loop Quantum Gravity like the semi-classical limit and the emergence of the space-time continuum among other things.
We want here to take advantage of the identification of the intertwiner space with the fuzzy sphere to show such a link.

The {\it fuzzy sphere}\cite{madore} is one of the best known example of NCG. The main idea of NCG is to describe geometry in terms of algebraic tools. Indeed a topological space $\mm$ is completely described by the commutative algebra $\cc^0(\mm)$ of continuous functions over it. If $\mm$ is further provided with a differential structure, to make it a manifold, $\mm$ gets encoded at the algebraic level as the algebra $\cc^{\infty}(\mm)$ provided with a differential operator. More precisely Connes \cite{connes} showed that there is a complete
equivalence between a spinorial compact manifold and a Spectral Triple $(\cc^{\infty}(\mm), \dd, \Gamma)$, which
contains the algebra, the Dirac operator encoding the differential structure and the metric, and finally the chirality operator. There is a list of axioms which constraint the possible choices of Spectral Triples, but we will not deal here with the details of the spectral triple structures. Nevertheless we will still call the differential operator as the Dirac operator.

There is a complete dictionary relating geometrical objects to algebraic ones. For example the diffeomorphisms of $\mm$ are encoded at the algebraic level by the automorphisms of $\cc^{\infty}(\mm)$. This construction is very fruitful as it allows in particular to unify Gravity with any Yang-Mills-Higgs theory as a Gravity theory on an almost-commutative space\cite{schucker}.

\medskip

Let us remind the mathematics of the fuzzy sphere. One considers the algebra of continuous functions of the sphere $\cc(\ss^2)$. These functions $f$ are generated by the polynomials $f(x)=f^0+f_ax^a+f_{ab}x^ax^b+\hdots$, together with the condition $x^ax_a=1$. The $x^a$ are the 3d euclidian coordinates and are naturally commutative. The fuzzy sphere is obtained by truncating the algebra in the degree of the polynomials. At the 0th order, the algebra is just $\C$. At first order, the vector space of linear polynomials is isomorphic to $\C^4$. As an algebra, one describe it as the matrix algebra $M_2(\C)$. This algebra is non-commutative and induces a non-commutative product on the space of polynomials. The interested reader can find details on the $\star$-product of the fuzzy sphere in \cite{freidel}.
One follows the same procedure at all orders: the fuzzy sphere $\ss^2_N$ is therefore described as the
matrix algebra $M_N(\C)$.

The coordinates for the fuzzy sphere are define as operators and are given by the generators of $\su(2)$, in the representation of spin $j=N/2$ (of dimension $(N+1)$):
\be
\label{coord}
X_i=\frac{1}{\sqrt{\frac{N}{2}(\frac{N}{2}+1)}}J_i^{(N/2)}.
\ee
They are non-commutative:
$$
[X_i, X_j]=i\frac{1}{\sqrt{\frac{N}{2}(\frac{N}{2}+1)}}\epsilon_{ij}{}^k\,X_k.
$$
This commutator vanishes in the limit $N\dr \infty$ and one recovers the usual classical sphere $\ss^2$. The "diffeomorphisms" of $\ss^2_N$ are given by the group of automorphisms of $M_N(\C)$, which is $\U(N)$.

Now, it is easy to make the link between the intertwiner space and the fuzzy sphere. Indeed, the space $\u(N)$ of $\SU(2)$ gauge invariant operators is naturally represented on the Hilbert space $M_N(\C)$ of the fuzzy sphere\footnote{One can also see $\u(N)$ as the space of Hermitian operators while $M_N(\C)$ is the algebra of all operators.}. Then the group $\U(N)$ generated by the intertwiner algebra is automatically identified with the diffeomorphisms of $\ss^2_N$. 

The geometry and metric information of the fuzzy sphere is encoded in the Dirac operator $\dd_N$. This operator acts on the algebra as a derivation, through the adjoint action: $a\in M_N(\C)\mapsto[\dd_N, a]$. Acting with $\dd_N$ therefore generates the 1-forms and allows to define the differential calculus on the fuzzy sphere. Moreover the Dirac operator allows to define a distance and therefore the metric \cite{connes, pierre}. On the algebra
$M_N(\C)$, a generic Dirac operator, or derivation operator, is simply an Hermitian operator and will be given as a linear combination of the $\u(N)$ generators, that is the intertwining operators $E_i$ and $G_{ij},G_{ij}^\dagger$. That is the Dirac operator defines a particular intertwiner. We can also see the Dirac operator as defining a particular basis of the intertwiner space as the basis of the eigenvectors of the derivation map $i[\dd_N,\cdot]$.
The fuzzy Dirac operator is defined in terms of an arbitrary vector $v^a $ and fuzzy coordinates:
$$
\dd_N=v^aX_a.
$$
In the limit $N\dr \infty$, we recover then the usual differential operator $\dd = v^a \epsilon_{ab}{}^cx^b\partial_c$  on the the classical sphere $\ss^2$ \cite{madore}.

\medskip

A natural question is what happens when we choose another Dirac operator: would we be simply describing a deformation of the fuzzy sphere? It is then important to point out that the matrix algebra $M_N(\C)$ is the algebra for any fuzzy surface. It is only when specifying the Dirac operator (or equivalently the coordinates) that we specify the topology and geometry of the considered fuzzy surface and define the classical surface we will get in the classical limit $N\dr\infty$. Indeed instead of taking the fuzzy sphere coordinates (\ref{coord}), one could have taken  the non-commutative torus coordinates $(u,v)$ defined as:
\be
u^n=1, \; v^n=1, \; uv=e^{i\frac{2\pi}{N}}vu.
\ee
Of course, the $N\dr\infty$ limit of this fuzzy torus leads back to the classical torus. More generally, it was conjectured by Madore \cite{madore2} that all the other genera are also given in terms of a specific basis in $M_N(\C)$. This means that before the choice of the Dirac operator, one is dealing with an arbitrary fuzzy surface. The choice of the Dirac specifies the topology that we recover in the limit $N\dr\infty$. $M_N(\C)$
contains very few information by itself, and it is the Dirac operator which contains the whole geometrical information. The eigenvectors of this Dirac operator \footnotemark  defines a preferred basis of the Hilbert space of intertwiners. This provides us with the geometrical interpretation of intertwiners: specific intertwiners will correspond to specific fuzzy surfaces. These intertwiners will be the ones describing the corresponding classical surface in the limit of an infinite number of punctures $N$.

\footnotetext{One of the important features of the Dirac operator in non-commutative geometry is actually its eigenvalues. In the context of the Spectral Geometry, one can derive from them all the information on the manifold starting from its dimension, or a simple counting of these eigenvalues allows to define the Spectral action which gives back the Einstein-Hilbert-Yang-Mills-Higgs action of quantum field theory coupled to general relativity in the almost-commutative geometry case \cite{connescham}.}


We see that non-commutative geometry provides us with a new point of view to study the (semi-)classical limit of loop quantum gravity. Defining classical surfaces as the limit of matrix algebras of increasing size, we are able to identify the exact geometrical meaning of intertwiners using the notion of Dirac operator. More generally, the mathematics of these matrix algebra limits is well understood and described in terms of Bratelli diagrams \cite{AF}. This is a path to explore, which we leave for future investigation.


Our last comment is about {\it Decoherence-Free Subspaces} (DFS). These spaces encodes the degrees of freedom which are not compromised by some chosen interactions or decoherence effects. These are especially important in the context of Quantum Computing where they identify the qbits which are most protected from decoherence effects. Following an idea by David Poulin and Fotini Markopoulou, \cite{lqgbh} proposes to identify the classical degrees of freedom
of a quantum black hole in the continuum limit in a particular DFS, which would carry a representation of the classical $\SU(2)$ rotation symmetry group. Our construction can be transcribed in that context. The space of intertwiners can be interpreted as the DFS of loop quantum gravity boundary degrees of freedom invariant under global $\SU(2)$ gauge transformation on the boundary. In this DFS, we have described the states which can be understood as the fuzzy sphere and thus carry an approximation of the classical symmetry group $\SU(2)$.


\subsection{LQG Dynamics from Matrix Model}

Let us consider the general set-up of a bounded region (with the simple topology of a 2-ball) of the three-dimensional space. At the quantum level, the interior of the region is described by a spin network state or more generally by a superposition of spin network states. Let us assume that the graph of all considered spin networks intersect the boundary surface in $N$ punctures. Then the interior geometry is described by some spin network with support on some graph (possibly very complex) as long as it only intersect the boundary surface in $N$ points.  The interior can be described by a superposition of spin networks with support on different graphs as long as all these graphs puncture the boundary sphere only at $N$ points. 
More generally, we could (or should) consider that the boundary surface be described by a superposition of states with different number of punctures. However, we then face the issue of defining the surface itself. Here we take the point of view that the surface is defined as the set of $N$ points. The only assumption is that it is then possible to do a splitting between what is the inside or outside this discrete surface made out of these $N$ points. Again in general, one would expect that such a splitting holds only approximately and becomes clear only at a coarse-grained level.

Let us thus consider a $N$-punctured 2-sphere. The state of the interior can be arbitrarily complicated, but from the boundary point of view, we only care about the resulting intertwiner between the $N$ $\SU(2)$ representations living at the punctures. The state of the  interior can involve a superposition of spin networks with support on different graphs, and it would not change as far as the boundary state is concerned since only the resulting intertwiner matters and not the details of the graphs. 
This induced boundary state can be described with a single vertex inside the 2-ball and can thus be interpreted as a completely coarse-grained bulk state: while the bulk could be described with complicated graphs, we can coarse-grain all these graphs to a simple graph with a single vertex as long as the intertwiner between the $N$ punctures is left unchanged, and an observer looking at the boundary will not see the difference. Having realized that the boundary state is simply a coarse-grained bulk state, we can propose a "holographic principle" stating that the dynamics of the boundary state can be described without referring to the full bulk state, since it is simply the dynamics of a single intertwiner in LQG. From this point of view, it does make sense to talk about the dynamics of the boundary state.

\begin{figure}[t]
\begin{center}
\includegraphics[width=8cm]{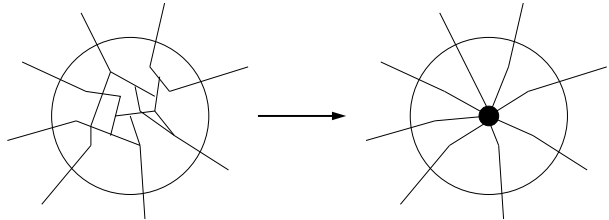}
\end{center}
\caption{Considering a bounded region of space, its boundary defined as a $N$ punctured sphere $\ss^2$, the LQG bulk degrees of freedom are described as all spin networks states with support on arbitrary graphs puncturing the surface $N$ times. Then one can coarse-grain these states down to a single vertex: the state is then described by a single intertwiner and is interpreted as the boundary state.}
\end{figure}

\medskip

Now, the operators $E,G$ define the gauge invariant degrees of freedom leaving on the surface once its area is fixed. Since these operators form a $\u(N)$ algebra, we can describe the degrees of freedom of the boundary surface as a $N\times N$ matrix and define the dynamics of the surface state through a {\it matrix model}. This matrix $M$ is the configuration variable on which act all the operators. Its diagonal entries correspond to classical values of the $E_i$ and its off-diagonal components to classical values of the $G_{ij}$. The matrix model would be defined through an action $S[M]$. Requiring the theory to be diffeomorphism invariant means asking the action to be invariant under the $\U(N)$ action $M\arr UMU^{-1}$. Therefore, the possible actions are generated by the traces of the matrix $Tr(M^k)$ or equivalently by symmetric polynomials of the eigenvalues of $M$. The eigenvalues of $M$ can be interpreted as proper vibration modes of the boundary surface.

Finally, as the matrix represents the degrees of freedom of the boundary, one can interpret this $U(N)$ gauged matrix model as the holographic theory for the bulk Loop Quantum Gravity. Indeed, any dynamics of the bulk spin network will simply induce a $\SU(2)$ gauge invariant dynamics on the boundary and thus will be described by some $\U(N)$ transformations. This looks very similar to the set-up of string theory where one works with $\U(N)$ super Yang-Mills theories on the boundary as the holographic theory for supergravities.

\medskip

More work is needed to propose a reasonable matrix model describing the LQG dynamics. Nevertheless, this matrix model formalism is very similar to the physics of 2d quantum gravity or higher-dimensional spin foam models\cite{carlogft} (group field theory formulation) and to the matrix model approach to string theory\cite{taylor}. This reformulation of LQG could therefore be a first explicit step towards making a bridge between the physics of LQG and the physics of M-theory (following the hopes of \cite{lee}), or towards writing a string theory on a spin network background.

%
%

\section{About Dynamics on a Spin Network}

The most pressing issue in LQG is understanding the dynamics of the quantum states of (3d) geometry. This can be interpreted as solving the Hamiltonian constraint for General Relativity (at the quantum level) and obtain the Hilbert space of physical states. Nevertheless, even then, the problem would not be solved. One should then understand the different physical choices of time, their corresponding gauge fixing and their corresponding Hamiltonian dictating the time evolution of the physical states.

The point that we would like to emphasize here is that these physical states are supposed to be spin network states (or precisely a superposition or distribution over spin networks). However, these states representing a 3d geometry would then be supposed to represent a space-time structure. It is thus a natural question to ask how can a spin network represent a dynamical process (happening in spacetime). We propose to view a spin network as a static space-time structure, on which information can flow along the links of the graph. From the spin foam point of view, we can point the spin network as the simple spin foam where the spin network is trivially evolved in such a way that there is no spacetime vertices. While the spin network describes a static spacetime set-up, one can talk about the dynamics of degrees of freedom living on the spin network. Roughly the considered degrees of freedom are the vectors in the $\SU(2)$ representations living on the edges, they evolve along the edges then meet and intertwine at the vertices of the graph. Starting with a set of vectors on some edges, one can follow their evolution through the spin network and see how they are processed. This way, a spin network state is interpreted as a {\it quantum circuit}: the 3d quantum space is identified with the set of processes that can occur (on it) or equivalently with the set of channels along which the (quantum) information can flow. We try to make this concept more precise in the next paragraphs.

\medskip

Working with $\SU(2)$, it is possible to decompose all representations into a (symmtrized/antisymmetrized) tensor product of the fundamental spin $1/2$ representation. A spin $1/2$ representation is called a {\it qubit} in the field of quantum information and quantum computing. These qubits meet and are processed at the intertwiners of the spin network, which can be precisely viewed as a circuit. The fundamental element of these circuits are 4-valent vertices of the spin network which can be interpreted as 2-qubits gates by considering two of the spin $1/2$ representations as ingoing and the two others as outgoing. The two fundamental intertwiners are the projection on the singlet (spin 0) states and the projection on the spin 1 states. Indeed, noting $V_j$ the spin $j$ representation, we have
$$
V_{1/2}\otimes V_{1/2}=V_0\oplus V_1,
$$
and we get define the orthogonal projections $P_0$ of rank one and $P_1$ of rank three respectively projecting on $V_0$ and $V_1$. $P_0$ is the projection on the Bell state $\phi_-=|\upa\downa\ra-|\downa\upa\ra$ while $P_1$ projects onto the three other Bell states $\phi_+=|\upa\downa\ra+|\downa\upa\ra$ and $\psi_\pm=|\downa\downa\ra\pm|\upa\upa\ra$.
All possible intertwiners are linear combination of these two operators. A special class of operators are the unitary processes $e^{i\alpha}P_0+e^{i\beta}P_1$. Considering that the intertwiners describe all processes which are invariant under global $\SU(2)$ rotation of the two qubits, allowing all individual $\SU(2)$ rotations on each qubit allows us to write any two-qubit gate: we have an exact equivalence between spin networks and qubit circuits.

\begin{figure}[t]
\begin{center}
\psfrag{h}{$\f{1}{2}$}
\psfrag{a}{$0$}
\psfrag{b}{$1$}
\includegraphics[width=7cm]{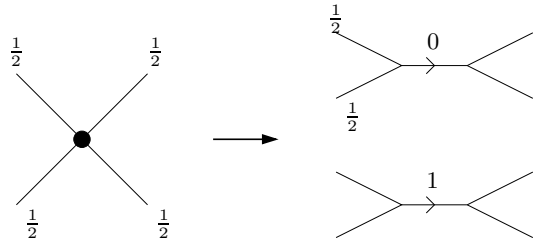}
\end{center}
\caption{Four-valent intertwiners between spin $1/2$ representations can be considered as the basic building block of spin networks. They are generated by two states, of spin $j=0$ and of spin $j=1$, and can be interpreted as two-qubit gates of a quantum circuit.}
\end{figure}

\medskip

Instead of decomposing the $\SU(2)$ representations into the fundamental spin $1/2$ representation, one can use the harmonic oscillator presentation and replace each edge of the spin network by a double edge, one line representing the "left-moving" oscillator $a$ and the other representing the "right-moving" oscillator $b$. These lines meet at the vertices at the graph where they get intertwined. In this formulation, a choice of intertwiner is a choice of (linear combination of) $\u(N)$ operators $E_i,G_{ij}$ acting on the double lines arriving at the vertex. Note that the $G_{ij}$ simply acts by moving a quanta of "energy" from the oscillators $a_i,b_i$ to the oscillators $a_j,b_j$, so that the action of intertwiners is rather simple. The dynamics on a spin network concerns the flow of the quanta levels of the harmonic oscillators, or more precisely the intertwiners defining the spin network define the evolution of the quanta levels of the harmonic oscillators along the graph underlying the spin network.

Now one would really describe the dynamics of the spin network states. We can think of two ways to possibly implement a Hamiltonian constraint or a Hamiltonian evolution operator. One could write the evolution as acting on the harmonic oscillators, not necessarily as the usual standard Hamiltonian. Or one could write the evolution as acting on the intertwiners, evolving the $\u(N)$ states living at the vertices. This latter description would be a generalization to a full spin network of the matrix model dynamics for an intertwiner. Nevertheless, naively, these Hamiltonian operators would act on spin network states on a particular graph. Either, one has in mind an effective theory where the graph only describes the 3d space at a coarse-grained level, with the vertices representing the spacetime points or chunk of 3d volumes as seen by an observer. Or one should go further and describe a background-independent  Hamiltonian operator possibly acting on spin networks with support on arbitrary graphs.


\section{The Geometry of a Spin Network and Quantum Reference Frames}

Although the area and volume operators have been implemented in Loop Quantum Gravity and are well-understood, one can not yet say that the geometry of spin networks is perfectly understood even at the kinematical level. Indeed there are two important issues actually related to the (space) diffeomorphism invariance of the quantum states of the theory. A first question is how to explicitly localize a system in a background independent formalism. This issue needs to be formulated in relational terms: one localizes an object with respect to another one. A second question is how to make sense of a superposition of spin networks, or in more precise words, where is one spin network compared to another one if there are both diffeomorphism invariant objects. These two topics are in fact closely related and the recurring theme is the notion of "where" on a spin network state.

\medskip

To make sense the location of a spin network with respect to another one in a quantum superposition, one should introduce objects, like matter degrees of freedom, on the spin network which can allow to pinpoint locations on that spin network. One can also introduce by hand geometrical objects like surfaces. From this point of view, a set-up which avoids the conceptual difficulties linked to the interpretation of a superposition of a spin network is the simple configuration of a region bounded by a 2-surface. Once the surface is defined as a set of a fixed number of punctures, one can have any complicated superposition of spin networks describing the geometry of the internal region, but as far as the boundary state is concerned, only the resulting intertwiner between the punctures is relevant and the details of the graphs defining the interior do not matter.

This viewpoint can be generalized to any (open or closed) 2-surface. Let us consider a surface in space, as made as $N$ discrete patches or $N$ punctures. Any spin network state (possibly superpositions) for the space geometry will puncture the considered surface and induce an intertwiner between the $N$ punctures: the intertwiner defines the quantum state of the surface and describes how the surface is folded i.e. how it is curved and embedded into the 3d space. This way, the surface state can be defined as a set of probability amplitudes for all possible intertwiners between the $N$ punctures and can be interpreted as the state of space geometry as seen by the surface. Now one can generalize these considerations to allow arbitrary numbers of punctures on the surface. However, our point of view is that the geometry state is viewed by some observer with some given resolution. He will effectively view the surface as made of a finite number of discrete patches (possibly cutting it into a lattice) and will describe the state of the surface at that resolution. Later, he get refine or coarse-grain his knowledge of the geometry of the space and of the surface by cutting the surface into more or less patches. Refining his knowledge of the surface state, the observer describes through a bigger number of patches or punctures. Thus from an effective viewpoint, it makes sense to fix the number of punctures $N$ to describe a given surface. But then an important question is how does the surface state (or space geometry state as seen by the surface) gets "renormalized" under refining or coarse-graining. It is obvious that the mean values of the geometrical quantities will change\footnote{For the example, the area of a surface will change except if it's flat. In a spherical space, the area of surface will increase under refinement while it would decrease in a hyperbolic space. So the renormalisation of geometrical quantities is relevant to the understanding of the space geometry.}.

A more complete and perhaps satisfying definition of a surface, taking into account the physical relevance of the number of punctures and the possibility of superpositions of states with different numbers of punctures, should involve matter degrees of freedom in order to localize what we call the surface. Else, in the pure geometry theory, nothing tells you that you is still talking about the same surface when changing the number of punctures. So one needs to first define the surface (with given number of punctures) and only then talk about the surrounding geometry state. This is equivalent to involving an observer who chooses the surface he wants to observe.

\medskip

Pushing the problem of localizing and defining a particular surface in a diffeomorphism invariant theory, one can say that one can only talk about one unique surface of the 2-sphere topology (up to possible windings in the 3d space manifold) since all similar surfaces can be mapped into each other by diffeomorphisms. Considering the abstract object of a closed surface with $N$ punctures, a quantum state of geometry then gives the quantum state of the surface i.e. the probability amplitude for all possible intertwiners between the $N$ punctures.  Turning the argument upside down, one could decide to define the quantum geometry state as the probability amplitudes for intertwiners for all (abstract) closed surfaces defined by their number of punctures $N$, their topology and their windings in the space manifold (depending on its topology) and all the correlations between the intertwiner probability amplitude distributions for two surfaces and three surfaces and so on \footnote{A priori, one should not only work with closed surface but should deal with open surfaces and generic tensor products instead of intertwiners.}. Such a definition of the quantum geometry state is similar to the Wick theorem in Quantum Field Theory which allows to reconstruct the Hilbert space and quantum vacuum state from all the correlations of the field(s). Finally, it would be very interesting to check whether there is actually an equivalence between the usual Hilbert space of states of spin networks of LQG and this definition of the quantum state of space geometry as the states of all classes of diffeomorphism equivalent surfaces (and the correlations between such surfaces).

\medskip

Now, let us look more closely at the issue of localizing an object on a spin network state (or on a superposition). In a diffeomorphism invariant theory, only relations (correlations) between objects have  physical meaning and the theory has to be understood in purely relational terms.  In this context, the notion of quantum reference frame becomes relevant. Indeed, when all is available are correlations between two subsystems $A$ and $B$, an interesting physical interpretation comes when $A$ can be interpreted as an approximate reference frame, i.e. that it is a system whose state space approximates the group algebra of a classical symmetry group. Particularly relevant cases are the $\U(1)$ group when $A$ defines a quantum clock, or $\C$ to describe a measurement, or any subgroup of the diffeomorphism in order to get a quantum reference frame allowing to localize objects in a quantum geometry.
In the previous sections, we have identified a system - a surface with the sphere topology defined as $N$ punctures with arbitrary spins - whose state space is group algebra of $\U(N)$ which can be interpreted as an approximation of the diffeomorphism group of a sphere: we have built a (quantum) reference frame allowing to localize points/systems/objects on a sphere. It would be interesting to generalize this one-sphere system to a multi-sphere system (each sphere inside the next one) which would then allow to localize directions in the 3d quantum space.

Pushing further the relational viewpoint, one can entirely view the quantum space state as defining a set of correlations/entanglements between the subsystems of the spin networks or equivalently regions of space, the elementary subsystems of a spin network being the nodes and corresponding intertwiners of the spin network (or alternatively one could choose the edges and vectors living on them). Then one could forget about the underlying spin networks and simply describe the geometry state as the correlations between a certain number of subsystems representing some regions of space. In such a set-up, there is a natural notion of distance between two objects. Indeed, in a diffeomorphism invariant framework, it makes sense that two objects are said to be close to each other when there are highly correlated. We thus propose to construct a notion of distance between two regions of a spin network state (possibly superposition) from their degree of entanglement (or more precisely the ration of their actual entanglement to their maximally allowed entanglement). Such construction is to be thought as similar to the entanglement or correlation laws derived in condensed matter systems such as spin systems where the entanglement decreases monotonously with the distance. This will be investigated in more details in \cite{distance} when we will try compute correlations between the harmonic oscillators or qubits on some spin network background.

\section{Outlook}

To conclude, this work is only preliminary, but the reformulation of LQG in terms of harmonic oscillators helps understanding quantum states of surfaces and seems promising both for the study of the semi-classical states or the dynamics and as a mathematical tool to probe the structure of the spin networks.
More precisely, we can identify the holographic (kinematical) degrees of loop quantum gravity and formulate their dynamics in terms of matrix model. This creates an explicit link between LQG and non-commutative geometry and provides us with new tools to analyze the continuum limit of spin networks and how they can describe the usual smooth manifolds in that limit.  Moreover it seems to also open a gate from LQG to string theory.
There are many possible extensions to the present work:

\begin{itemize}

\item From the point of view of LQG, it would be interesting to build the operators corresponding to the holonomies, in order to get the full LQG algebra. On the spin network basis, these operators correspond to make the tensor product with some given $\SU(2)$ representation of all the representations on the edges of a given loop of the spin network. This would allow a full reformulation of LQG.
\item It would be interesting to generalize our intertwiner analysis to all surface topologies. This would mean considering a vertex with $N$ external legs and a given number $g$ of loops which correspond to the non-contractible cycles of the surface. Then one would like to impose the holonomy around these loops. This would be helpful to understand the quantum states of (closed) surfaces of higher genus and also in order to explore the bridge with string theory. For the point of view of non-commutative geometry and fuzzy surfaces, we should always find the same $\u(N)$ algebra. Nevertheless different intertwiners, or equivalently  $\u(N)$ states, will be interpreted as fuzzy surfaces of different topologies and the interpretation becomes clearer in a continuum limit $N\arr\infty$.
\item One could investigate the extension of the formalism to super-gravity. For $N=1$ super-gravity, we would have spin networks with representations of $\lalg{osp}(1|2)$, and it turns out possible to present $\lalg{osp}(1|2)$ in terms of harmonic oscillators as we show in appendix.
\item Since the harmonic oscillator presentation of $\su(2)$ canonically provides a representation of the Lorentz algebra $\sl(2,\C)$ (see in appendix), it would be interesting to identify the resulting Lorentz invariant states and check whether they have any relevant physical meaning.
\item One can generalize our construction to spin foam models a la Barrett-Crane\cite{BC}.  Then spin networks are labeled by simple representation of the Lorentz group ($\SO(4)$ in the Riemannian version and $\SO(3,1)$ in the Lorentzian version). We can present the Lorentz algebra as a couple of left/right $\su(2)$ algebra and thus as a couple of couple of harmonic oscillators $(a_L,b_L,a_R,b_R)$. The energy of the left harmonic oscillators define the left spin while the energy of the right sector defines the right spin. Then simple representations are obtained by imposing the constraint that the energies of the right and left sectors are equal.
\item We would like to study more in details the construction of quantum reference frames in spin networks. This creates a strong link with quantum information where people also try to construct reference frames out of sets of qbits and study their robustness and transition to classicality. This means using some tools developed in the quantum information field to tackle some of the quantum gravity problems. But  we can also export the quantum gravity results on spin networks to quantum information, using in general the equivalence between quantum circuits and spin networks. On a more precise level, we have presented a system made of $2N$ harmonic oscillators, or equivalently $N$ systems of arbitrary spins, and showed that the decoherence-free subspace under global $\SU(2)$ errors is $\U(N)$. This could be interesting to realize or code unitary operators in quantum computing, or to transport unitaries along a quantum channel while ignoring the reference frames used by the two parties. More generally, we would like to explore how the geometrical information -distance and curvature- gets encoded in terms of information, entropy and entanglement in the context of loop quantum gravity.

\end{itemize}

And finally, the most interesting and difficult problem is the issue of the dynamics: understanding which dynamics to give to the intertwiners and matrix models for boundary states. We could draw some inspiration from the spectral actions of non-commutative geometry (using the Dirac operator) or from the matrix models of string theory.

\section*{Acknowledgements}
We would like to thank Laurent Freidel, Carlo Rovelli, Lee Smolin for their enthusiasm and encouragements in completing this work, and Danny Terno, Rob Spekkens and Paolo Zanardi for their interest into this work from the point of view of quantum information. We are grateful to Daniele Oriti for many discussions on the interplay between Quantum Information and Loop Quantum Gravity. We are particularly thankful to Simone Speziale for a careful reading of an early draft and his comments. EL would also like to thank Stephon Alexander for a discussion on matrix models in string theory and Harold Ollivier for a discussion on two-qubits gates.

\appendix

\section{Harmonic Oscillator presentation of $\osp(1|2)$}

Considering an extra fermionic harmonic oscillator:
$$
\{c,c\dag\}=1, \quad c^2=(c\dag)^2=0,
$$
one can define the following operators:
\be
Q_+=\f{1}{2}\left(a\dag c+c\dag b\right),\quad
Q_-=\f{1}{2}\left(a c\dag-c b\dag\right).
\ee
These operators satisfy the following commutation relations:
$$
\{Q_+,Q_-\}=\f{1}{2}J_z,\quad
\{Q_\pm,Q_\pm\}=\pm\f{1}{2}J_\pm,
$$
$$
[J_\pm,Q_\pm]=0,\quad
[J_\pm,Q_\mp]=-Q_\pm,
$$
$$
[J_z,Q_\pm]=\pm\f{1}{2}Q_\pm,
$$
so that the operators $\v{J},Q_\pm$ form the $\osp(1|2)$ supersymmetric Lie algebra. One can check that the invariant Casimir operator is ${\cal C}=J^2-{\cal C}_f$ where:
$$
{\cal C}_f=Q_+Q_--Q_-Q_+
=\f{1}{2}\left(E[c,c\dag]-c\dag c\right)=\f{1}{2}\left(E-c\dag c(1+2E)\right).
$$

The Hilbert space for the fermionic harmonic oscillator has of course only two states:
$$
c|+\ra=|-\ra, \quad
c\dag|-\ra= |+\ra,
$$
so that the energy $c\dag c$ takes the value 0 (on $|-\ra$) or 1 (on $|+\ra$). For a given bosonic energy level $E=j$, the fermionic part of the Casimir operator takes two possible values:
$$
{\cal C}_f\,|j,-\ra=\f{1}{2}j|j,-\ra,\quad
{\cal C}_f\,|j,+\ra=-\f{1}{2}(j+1)|j,+\ra,
$$
where we have ignored the magnetic moment $m$.
Then the vector space of the supersymmetric representation of spin $j$ for $\osp(1|2)$ consists of the vectors
$|j,-\ra$ and $|j-1/2,+\ra$. The corresponding Casimir is ${\cal C}=j(j+1/2)$. For more details on the representation theory of $\osp(1|2)$ and the corresponding spin networks, the reader can look at \cite{susysf}. 

\section{A canonical representation of the Lorentz algebra $\sl(2,\C)$}

Starting from the presentation of $\su(2)$ in terms of harmonic oscillators, one can naturally introduce a representation of the Lorentz algebra, which includes all the representations of $\su(2)$. We introduce boost generators:
\be
K_z=\f{1}{2}\left(a\dag b\dag+ab\right),\quad
K_+=\f{1}{2}\left(b^2-(a\dag)^2\right),\quad
K_-=\f{1}{2}\left((b\dag)^2-a^2\right),
\ee
which then satisfy the commutation relations:
\be
\begin{array}{ll}
[J_z,K_z]=[J_\pm,K_\pm]=0, & {[}J_z,K_\pm{]}=\pm K_\pm, \\
{[}K_z,J_\pm{]}=\pm K_\pm, & [ K_z,K_\pm ]=\mp J_\pm, \\
{[}J_\pm,K_\mp{]}=\pm 2K_z, & [K_+,K_-]=-2J_z.
\end{array}
\ee
The Casimir operators of $\sl(2,\C)$ are:
$$
{\cal C}_1=J_z^2+\f{1}{2}\left(J_+J_-+J_-J_+\right)-K_z^2-\f{1}{2}\left(K_+K_-+K_-K_+\right),
$$
$$
{\cal C}_2=J_zK_z+\f{1}{2}\left(J_-K_++J_+K_-\right).
$$
It is easy to check that this harmonic oscillator representation of $\sl(2,\C)$ is a unitary representation with
${\cal C}_1=-3/4$ and ${\cal C}_2=0$. Unitary representations of $\sl(2,\C)$ are labeled by a couple of positive half-integer and real non-negative number $(n,\rho)$ (as far as the principal series are concerned). They decompose onto $\su(2)$ representations of spin $j\ge n$. The present harmonic oscillator of $\sl(2,\C)$ is the simple representation $(n=1/2,\rho=0)$.



\begin{thebibliography}{99}

\bibitem{lqg}
C Rovelli,
{\it Loop Quantum Gravity},
Living Rev.Rel. 1 (1998) 1,
gr-qc/9710008; \\
T Thiemann,
{\it Lectures on Loop Quantum Gravity},
Lect.Notes Phys. 631 (2003) 41-135,
gr-qc/0210094; \\
A Ashtekar, J Lewandowski,
{\it Background Independent Quantum Gravity: A Status Report},
Class.Quant.Grav. 21 (2004) R53,
gr-qc/0404018

\bibitem{marzuoli}
A Marzuoli, M Rasetti,
{\it Spin network quantum simulator},
Phys.Lett. A306 (2002) 79-87,
quant-ph/0209016; \\
A Marzuoli, M Rasetti,
{\it Spin network setting of topological quantum computation},
quant-ph/0407119; \\
A Marzuoli, M Rasetti,
{\it Computing Spin Networks},
quant-ph/0410105


\bibitem{screen}
F Markopoulou, L Smolin,
{\it Holography in a quantum spacetime},
hep-th/9910146

\bibitem{taylor}
W Taylor,
{\it The M(atrix) model of M-theory},
Lectures for NATO school "Quantum Geometry" (Iceland 1999),
hep-th/0002016

\bibitem{goodman}
R Goodman, NR Wallach,
{\it Representations and Invariants of the Classical groups}, 
Cambridge University Press (1998)

\bibitem{cohstates}
ER Livine, D Oriti, S Speziale,
{\it Coherent Quantum Tetrahedron for (Loop) Quantum Gravity}, in preparation

\bibitem{barbieri}
A Barbieri,
{\it Quantum tetrahedra and simplicial spin networks},
Nucl.Phys. B518 (1998) 714-728,
gr-qc/9707010


\bibitem{perelomov}
A Perelomov, {\it Generalized Coherent States and Their Applications},
Springer-Verlag (1986)

\bibitem{madore}
J Madore,
{\it The fuzzy sphere},
Class.Quant.Grav. 9 (1992) 69

\bibitem{connes}
A Connes,
{\it Noncommutative Geometry},
Academic Press (1994)


\bibitem{freidel}
L Freidel, K Krasnov,
{\it The fuzzy sphere $\star$-product and spinetworks},
J.Math.Phys. 43 (2002) 1737,
hep-th/0103070

\bibitem{schucker}
T Schucker,
{\it Forces from Connes geometry},
hep-th/0111236

\bibitem{pierre}
B Iochum, T Krajewski, P Martinetti,
{\it Distances in finite spaces from non commutative geometry},
J.Geom.Phys. 37 (2001) 100,
hep-th/9912217

\bibitem{madore2}
J Madore, LA Saeger,
{\it Topology at the Planck length},
gr-qc/9708053

\bibitem{connescham} 
A Chamseddine, A Connes,
{\it The spectral action principle},
Comm.Math.Phys. 186 (1997) 731,
 hep-th/9606001

\bibitem{AF}
G Landi,
{\it An introduction to noncommutative spaces and their geometry},
hep-th/9701078

\bibitem{lqgbh}
O Dreyer, F Markopoulou, L Smolin,
{\it Symmetry and entropy of black hole horizons},
hep-th/0409056



\bibitem{carlogft}
M Reisenberger, C Rovelli,
{\it Spacetime as a Feynman diagram: the connection formulation},
Class.Quant.Grav. 18 (2001) 121-140,
gr-qc/0002095



\bibitem{lee}
L Smolin,
{\it A candidate for a background independent formulation of M theory},
Phys.Rev. D62 (2000) 086001,
hep-th/9903166;
{\it The cubic matrix model and a duality between strings and loops},
hep-th/0006137

\bibitem{distance}
F Girellli, ER Livine, D Terno,
{\it Reconstructing Quantum Geometry from Quantum Information: Entanglement as a Measure of Distance},
in preparation

\bibitem{BC}
JW Barrett, L Crane,
{\it Relativistic spin networks and quantum gravity},
J.Math.Phys. 39 (1998) 3296-3302,
gr-qc/9709028; 
{\it A Lorentzian Signature Model for Quantum General Relativity},
Class.Quant.Grav. 17 (2000) 3101-3118,
gr-qc/9904025

\bibitem{susysf}
ER Livine, R Oeckl,
{\it Three-dimensional Quantum Supergravity and Supersymmetric Spin Foam Models},
Adv.Theor.Math.Phys. 7 (2004) 951-1001,
hep-th/0307251

\end{thebibliography}
\end{document}